\documentstyle[12pt,epsf]{article}
\def\draftversion{N}			    
\def\note[#1]#2{\message{(#1)}\if\draftversion Y{\noindent\em #2}\fi}
\newif\ifenglish\englishfalse		    
\if \draftversion Y
\fi
\def\half{{\scriptstyle{1\over2}}}  
\newcommand{\mc}{Monte Carlo}		    
\newcommand{\qcd}{{\sc qcd}}		    
\newcommand{\rge}{renormalization group equation}   
\newcommand{\BF}{$\beta$-function}	    
\newcommand{\DB}{\Delta\beta}		    
\newcommand{\F}{\Omega}			    
\newcommand{\f}{f}			    
\renewcommand{\d}{\delta}		    
\renewcommand{\b}{\beta}		    
\renewcommand{\c}{\beta'}		    
\newcommand{\pert}{{\hbox{\rm\tiny perturbative}}}  
\newcommand{\dof}{{\hbox{\rm\tiny dof}}}    

\newcommand{\flavour}{flavo\ifenglish u\fi r}
\newcommand{\Flavour}{Flavo\ifenglish u\fi r}
\newcommand{\ensure}{\ifenglish e\else i\fi nsure}

\def\spose#1{\hbox to 0pt{#1\hss}}
\def\ltapprox{\mathrel{\spose{\lower 3pt\hbox{$\mathchar"218$}}
 \raise 2.0pt\hbox{$\mathchar"13C$}}}
\def\gtapprox{\mathrel{\spose{\lower 3pt\hbox{$\mathchar"218$}}
 \raise 2.0pt\hbox{$\mathchar"13E$}}}
\def\inapprox{\mathrel{\spose{\lower 3pt\hbox{$\mathchar"218$}}
 \raise 2.0pt\hbox{$\mathchar"232$}}}

\begin{document}

\title{QCD $\beta$ Function with Two {\Flavour}s of Dynamical Wilson Fermions}

\author{K.~M.~Bitar, R.~G.~Edwards, U.~M.~Heller, and A.~D.~Kennedy \\[1ex]
  Supercomputer Computations Research Institute, \\
  Florida State University, Tallahassee, Florida 32306-4052, U.S.A.}
\date{FSU--SCRI--96--07 (Revised)\\[0.75ex] March 22, 1996}

\maketitle

\begin{abstract}
  \noindent We test the asymptotic scaling behavior of state-of-the-art
  simulations of QCD with two {\flavour}s of light Wilson fermions. This is done
  by matching $\pi$ and $\rho$ masses on lattices of size $16^3\times32$ and
  $8^3\times16$. We find that at $\beta=6/g^2=5.3$ matching is not possible over
  a range extending down to $\beta=3.5$. The large lattice data at $\beta=5.5$
  matches the small lattice values at $\beta=4.9(1)$ leading to a shift
  $\Delta\beta=0.6(1)$, considerably larger than the perturbative prediction of
  $0.45$. In both cases we conclude that the simulations are very far from the
  asymptotic scaling region.{\parfillskip=0pt\par}
\end{abstract}

\section{Introduction}

Assuming that a universal continuum limit of lattice {\qcd} exists, measurements
of some physical observable $\F$, such as a hadron mass, in a {\mc} computation
will not depend on the details of the lattice discretization if the lattice
spacing~$a$ is small enough. The value of the observable, $\F=a^\d
\f(g,\kappa)$, obtained by numerical calculation in lattice {\qcd} with Wilson
fermions, depends only on the parameters appearing in the lattice action: the
gauge coupling~$\beta=6/g^2$ and the hopping parameter~$\kappa$. The
dimensionless quantity $f$ has been related to the dimensionful observable $\F$
by multiplying by the appropriate power of the lattice spacing $a$. As we
approach the continuum limit $a\to0$ the fact that $\F$ is independent of $a$ is
expressed by the {\rge}
\begin{displaymath}
  a {d\F\over da} = a^\d \left( \d\f
    - \b {\partial\f\over\partial g}
    - \c {\partial\f\over\partial\kappa}
    \right) = 0
\end{displaymath}
where
\begin{displaymath}
  \b = - a {dg\over da}, \qquad \c = -a {d\kappa\over da}.
\end{displaymath}
The derivatives with respect to $a$ are taken at ``constant physics.'' As we
have two independent relevant parameters we require two independent physical
observables to use as renormalization conditions to fix these parameters; $\b$
and $\c$ should, of course, be independent of the choice of renormalization
conditions.

In a perturbative expansion of $\b$ about $g=0$ only the first two terms are
independent of the details of the lattice regularization scheme. When the higher
order terms are negligible one generally speaks of being in the {\it asymptotic
scaling\/} regime. The aim of this work is to determine, for the realistic case
of {\qcd} with two {\flavour}s of light Wilson fermions, how close present
simulations are to this asymptotic scaling regime.

In numerical simulations it is impractical to determine the {\BF}, which
correspond to an infinitesimal change of scale, directly. Rather we work with a
finite change of scale. We consider two lattices, $\Lambda_1$ with $L_1$ sites
in each direction and with lattice spacing $a_1$, and $\Lambda_2$ with $L_2$
sites in each direction and lattice spacing $a_2$. By fiat we assert that
$\Lambda_1$ and $\Lambda_2$ represent the same physical volume, thus $L_1 a_1 =
L_2 a_2$.

If we select some set of physical observables $\F_i$ we have
\begin{displaymath}
  \F_i = a_j^{\d_i} \f_i(g(a_j),\kappa(a_j)),
\end{displaymath}
where $\d_i$ is the dimension of $\F_i$ and $\f_i$ is the corresponding
dimensionless function measured on the lattice. In the continuum limit all these
observables must become independent of the lattice, so for the two different
lattices
\begin{displaymath}
  \f_i \Bigl( g(a_1), \kappa(a_1) \Bigr) a_1^{\d_i}
    = \f_i \Bigl( g(a_2), \kappa(a_2) \Bigr) a_2^{\d_i}.
\end{displaymath}

We define the quantities
\begin{displaymath}
  \DB \equiv {6\over g^2(a_2)} - {6\over g^2(a_1)},
    \qquad \Delta\kappa \equiv \kappa(a_2) - \kappa(a_1)
\end{displaymath}
to be the change in the couplings needed to compensate for this change in the
cutoff.

\section{The \BF}

We expect the {\BF} to become independent of~$\kappa$ in the continuum and
chiral limits.\footnote{If we use continuum perturbation theory with dimensional
regularization and the minimal subtraction renormalization scheme, then the BRS
identities {\ensure} that all divergences are logarithmic, and thus must be
independent of the quark mass. From this we may conclude that in the vicinity of
the chirally symmetric continuum limit any $\kappa$ dependence of the {\BF} must
arise from irrelevant lattice operators and be suppressed by some power of the
lattice spacing.}

The {\BF} at fixed $\kappa$ is defined by $\b(g) = -a\,dg/da$, and the two
universal terms from perturbation theory are $\b(g) = -b_0 g^3 - b_1 g^5 +
O(g^7)$ where
\begin{eqnarray*}
  b_0 &=& {33 - 2 n_f \over 48\pi^2} \\
  b_1 &=& {153 - 19 n_f \over 384\pi^4}
\end{eqnarray*}
for {\qcd} with $n_f$ {\flavour}s of light fermions.

The solution of these equations, expressed in terms of $\beta=6/g(a_2)^2$, is
\begin{displaymath}
  \ln\left({a_2\over a_1}\right)
    = {\sqrt6\over2} \int_{\beta-\Delta\beta}^\beta
      {dx\over x^{3/2} \b\left(\sqrt{6/x}\right)};
\end{displaymath}
after evaluating the integral we find that
\begin{displaymath}
  \Delta\beta_\pert \approx - 12 \left( b_0 + {6 b_1 \over \beta} \right)
    \ln\left( {a_2\over a_1} \right)
\end{displaymath}
which is the perturbative prediction for the change in $\beta$ needed to effect
a change in scale by a factor $a_2/a_1$.

$\DB$ has been thoroughly studied for pure gauge theory with the Wilson action
(see \cite{akemi93a} and references therein) and exhibits the well-known dip
around $\beta=6.0$. For {\qcd} with dynamical fermions we are unaware of recent
studies of $\DB$. For staggered fermions, Blum et al.~\cite{blum95a}, determined
the {\BF}, needed for the computation of the equation of state for finite
temperature {\qcd}, from fits to hadron masses obtained at different couplings,
and hence different lattice scales. They found significant deviations from
asymptotic scaling that again would lead to a dip in $\DB$.

\section{Results}

The {\BF} for two {\flavour}s of Wilson fermions was computed earlier by some of
us~\cite{bitar89b,kennedy89a}. The physical observables used for that purpose
were $m_\pi$ and several Creutz ratios of Wilson loops.

The main conclusion then was that perturbative scaling was only expected to
occur for $\beta>6.0$. In those simulations the $\pi$-like object was heavy, and
it was thought at the time that this may have hidden the effects of the
dynamical fermions in the system. Simulations with smaller $m_\pi$ are thus
needed to clarify the issue.

In this work we concentrated on data available on lattices of size
$16^3\times32$ with two {\flavour}s of Wilson fermions with parameters and
results shown in Table~\ref{table-one}~\cite{hemcgc93b,kennedy96b}. These were
matched using $8^3\times16$ lattices with the parameters and results for the
interesting parameter region shown in Table~\ref{table-two}.

\begin{table}
  \begin{center}
    \tabcolsep 4pt
    \begin{tabular}{ccccccccc}
      \hline
      $\beta$ & $\kappa$ & $\#$ & $\Delta$ & $am_\pi$ & $am_\rho$ & $am_N$
	      & $m_\pi/m_\rho$ & $m_\pi/m_N$ \\
      \hline
      5.3 & 0.1670 & 484 &  5 & 0.4540(20) & 0.6350(20) & 0.9620(40) &
	    0.715(5)  & 0.4719(29) \\
      5.3 & 0.1675 & 417 &  3 & 0.3120(40) & 0.5230(40) & 0.7660(90) &
	    0.596(15) & 0.4073(71) \\
      5.5 & 0.1596 & 400 &  5 & 0.3754(27) & 0.4947(78) & 0.7644(48) &
	    0.759(11) & 0.4911(36) \\
      5.5 & 0.1600 & 400 &  5 & 0.3262(17) & 0.4606(28) & 0.7106(91) &
	    0.708(4)  & 0.4591(57) \\
      5.5 & 0.1604 & 669 &  3 & 0.2666(20) & 0.4208(55) & 0.6326(56) &
	    0.634(8)  & 0.4215(48) \\
      \hline
    \end{tabular}
    \caption[table-one]{Parameters and masses on $16^3\times32$
      lattice~\cite{hemcgc93b,kennedy96b}. $\#$ is the number of measurements
      made, each separated by $\Delta$ trajectories.}
    \label{table-one}
  \end{center}
\end{table}

\begin{table}
  \begin{center}
    \tabcolsep 4pt
    \begin{tabular}{cccccccc}
      \hline
      $\beta$ & $\kappa$ & $\#$ & $am_\pi$ & $am_\rho$ & $am_N$
	      & $m_\pi/m_\rho$ & $m_\pi/m_N$ \\
      \hline
      4.8  & 0.1890 & 130 & 0.6681(41) & 0.964(12) & 0.964(12) &
	     0.678(16)  & 0.4038(51) \\
      4.8  & 0.1900 & 170 & 0.5814(35) & 0.897(7)  & 1.522(26) &
	     0.6480(54) & 0.3820(64) \\
      4.8  & 0.1905 & 109 & 0.5110(48) & 0.896(28) & 1.362(26) &
	     0.571(19)  & 0.3753(65) \\
      \hline
      4.9  & 0.1840 & 189 & 0.7826(28) & 1.013(5)  & 1.783(36) &
	     0.7722(36) & 0.4390(88) \\
      4.9  & 0.1845 & 190 & 0.7520(23) & 0.988(5)  & 1.714(20) &
	     0.7609(33) & 0.4388(49) \\
      4.9  & 0.1850 & 110 & 0.6808(46) & 0.928(5)  & 1.549(13) &
	     0.7334(44) & 0.4396(40) \\
      4.9  & 0.1855 & 272 & 0.6491(40) & 0.930(7)  & 1.660(57) &
	     0.6982(63) & 0.391(13)  \\
      4.9  & 0.1860 & 200 & 0.5636(47) & 0.872(10) & 1.451(31) &
	     0.6467(88) & 0.3886(88) \\
      4.9  & 0.1865 & 150 & 0.5018(59) & 0.852(20) & 1.55(11)  &
	     0.590(13)  & 0.324(22)  \\
      4.9  & 0.1870 & 70  & 0.4981(65) & 0.816(20) & 1.395(23) &
	     0.611(15)  & 0.3572(81) \\
      \hline
      4.95 & 0.1815 & 80  & 0.8083(37) & 1.003(6)  & 1.693(19) &
	     0.8059(44) & 0.4775(47) \\
      \hline
      5.0  & 0.1800 & 30  & 0.7734(85) & 0.990(14) & 1.690(36) &
	     0.781(11)  & 0.4578(95) \\
      5.0  & 0.1810 & 150 & 0.6321(53) & 0.875(7)  & 1.462(19) &
	     0.7228(53) & 0.4325(65) \\
      5.0  & 0.1815 & 270 & 0.5145(83) & 0.814(9)  & 1.342(19) &
	     0.632(12)  & 0.3834(77) \\
      \hline
    \end{tabular}
    \caption[table-two]{Parameters and masses for the matching region
      on $8^3\times16$ lattices. $\#$ is the number of measurements made.}
    \label{table-two}
  \end{center}
\end{table}

A preliminary report on this work was presented in~\cite{bitar94a}.

The values of $\beta$ were chosen for hadron spectrum and other calculations
largely based on the assumption that the presence of two fermion {\flavour}s
renormalizes $\beta$ downwards by about $0.5$ from its quenched value, and thus
we may not be far from the perturbative scaling regime at these parameter
values.

For our test of scaling we used two hadron masses, $m_\pi$ and $m_\rho$, as
observables for matching with similar data generated on lattices of size
$8^3\times16$ at values of $\beta$ ranging from $5.1$ down to $3.5$ and many
$\kappa$ values.

The $8^3\time16$ simulations used step sizes from $0.015$ to $0.025$ to
{\ensure} a reasonable acceptance rate. The conjugate gradient residual used for
the molecular dynamics steps was $10^{-6}$. Hadron measurements were made every
$5$ trajectories. We found integrated autocorrelation times were typically of
this order.

Quark propagators with wall sources and point sinks were computed in Coulomb
gauge on time slices $0$ and $L_t/2$. We determined the meson masses by a
correlated single state cosh fit to two correlation functions with interpolating
fields $\bar\psi \Gamma_i \psi$ and $\bar\psi \gamma_0 \Gamma_i \psi$; our
baryon correlation functions used only one interpolating field. The covariance
and errors were determined via the bootstrap method. Fits were made through the
center of the lattice with a starting time slice determined by the criterion
$\max{N_{\dof}Q\,\delta m/m}$, where $N_{\dof}$ is the number of degrees of
freedom for a fit of mass $m$. In all cases $Q$, the 1~standard deviation
confidence level, was greater than~$10\%$.

\subsection{Non-Matching at $\beta=5.3$}

\begin{figure}[htb]
  \epsfxsize=0.75\textwidth
  \centerline{\epsfbox{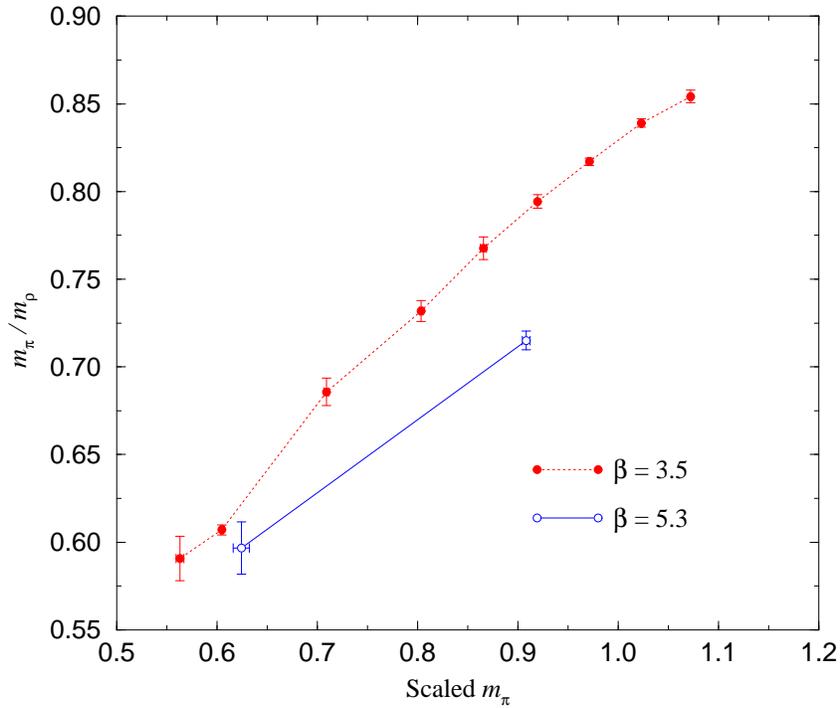}}
  \caption[fig1]{Non-matching of $m_\pi/m_\rho$ at $\beta=5.3$.}
  \label{fig1}
\end{figure}

Our first result is the unmatchability of the data at $\beta=5.3$. We find that
$\DB$ must be larger than $1.8$. Figure~\ref{fig1} shows the two points obtained
at $\beta=5.3$ on the larger lattice (with $m_\pi$ appropriately scaled by a
factor of 2) and the corresponding data for $m_\pi$ and $m_\pi/m_\rho$ on the
smaller volume for $\beta=3.5$. Data points obtained at larger $\beta$ values
lie to the left of those from $\beta=3.5$ (see figure~\ref{fig3}).

\subsection{Matching at $\beta=5.5$}

\begin{figure}[htb]
  \epsfxsize=0.75\textwidth
  \centerline{\epsfbox{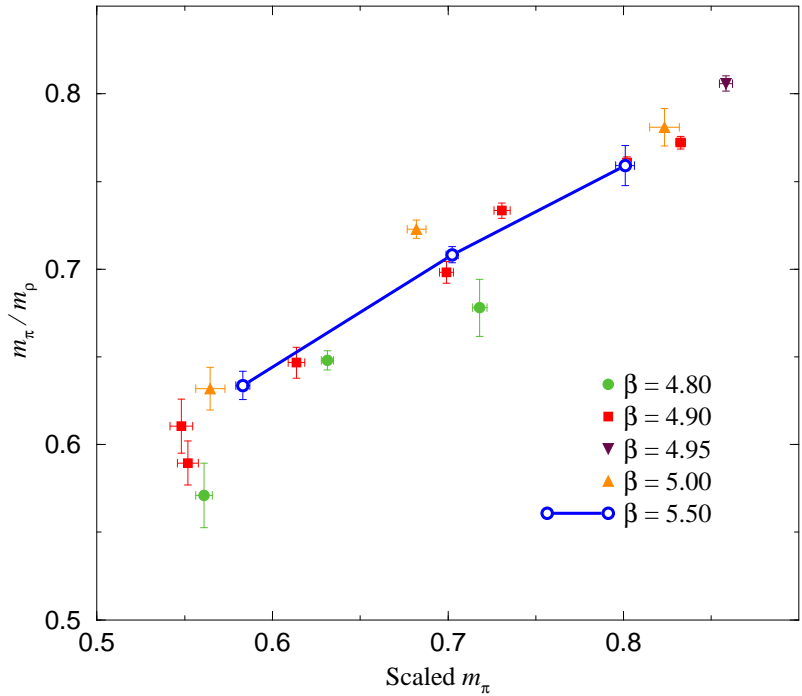}}
  \caption[fig2]{Matching of $m_\pi/m_\rho$ at $\beta=5.5$}
  \label{fig2}
\end{figure}

Our second result is that at $\beta=5.5$ matching is possible. Figure~\ref{fig2}
shows the  matching with $\beta=4.9$ on the smaller lattice. It is interesting
to note that all three points, at different values of $\kappa$, match with
corresponding points at the same value of $\beta$. For comparison similar points
obtained at $\beta=4.8$ and $5.0$ on the smaller lattice are shown. For $\beta=
4.8$ $m_\pi/m_\rho$ tends to be lower than the values at $\beta=5.5$, whereas at
$\beta=5.0$ they are higher. Two conclusions follow from this data: the first is
that the data at $\beta=5.5$ is consistent with a value of $\Delta\beta=0.6(1)$,
and second that this matching is the same for all three different values of
$\kappa$ used, indicating a $\kappa$ independence of the \BF.

At the central matching $\beta$ value ($\beta=4.90$) we find $\Delta\kappa =
-0.0249(1)$, $-0.0253(2)$, and $-0.0258(2)$ for large lattice values $\kappa =
0.1596$, $0.1600$, and $0.1604$ respectively, thus giving an estimate of the
integrated $\c$-function.

\begin{figure}[htb]
  \epsfxsize=0.75\textwidth
  \centerline{\epsfbox{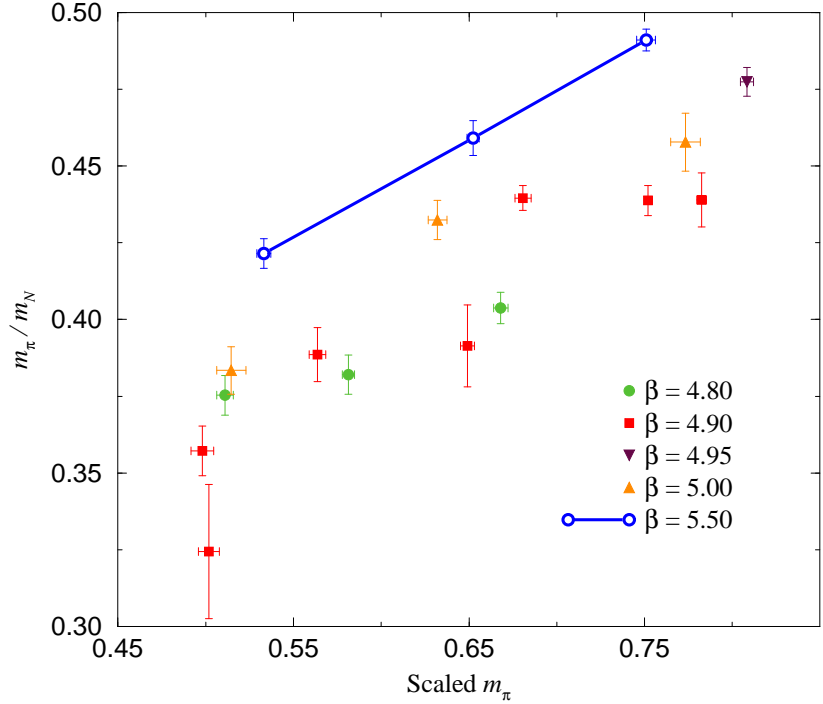}}
  \caption[fig2a]{Non-matching of $m_\pi/m_N$ at $\beta=5.5$}
  \label{fig2a}
\end{figure}

On the other hand, figure~\ref{fig2a} shows that the the system is not in the
scaling region because the pion-nucleon mass ratio does not match for the same
values of $\beta$ and~$\kappa$.

Another set of observables which we can match to determine $\DB$ are appropriate
ratios of Wilson loops, just as was done for pure gauge
theories~\cite{hasenfratz84b}. We computed space-like planar Wilson loops on all
our lattice configurations and evaluated ratios $R$ of two or four Wilson loops
arranged such that the corner and perimeter singularities cancel between
numerator and denominator,
\begin{eqnarray*}
  R(\beta, \kappa, \{r_1, r_2, r_3, r_4\}, L_s, L_t) &=&
    {W(r_1, r_2) \over W(r_3, r_4)},
    \qquad r_1+r_2 = r_3+r_4; \\
  R(\beta, \kappa, \{r_1, r_2, r_3, r_4, r_5, r_6, r_7, r_8\}, L_s, L_t) &=&
    {W(r_1, r_2) W(r_3, r_4) \over W(r_5, r_6) W(r_7, r_8)},
    \qquad \sum_{i=1}^4 r_i = \sum_{i=5}^8 r_i.
\end{eqnarray*}
We then determined $\DB$ from the condition
\begin{displaymath}
  R(\beta-\DB, \kappa', \{r\}, \half L_s, \half L_t) =
    R(\beta, \kappa, \{2r\}, L_s, L_t).
\end{displaymath}
$\kappa'$ on the smaller lattice was tuned such that the $\pi$ mass on the
smaller lattice was twice that on the larger lattice. In practice we
interpolated (or extrapolated) linearly from two simulations on the smaller
lattice with slightly different values of $\beta'$. Unfortunately, with our
statistics the errors on the larger Wilson loops grew quickly and they were not
useful for the ratio matching. With smaller Wilson loops dominating the ratios
lattice artefacts in the matching become important. We tried to avoid this by
considering tree level and one-loop improved ratios \cite{hasenfratz84b}. For
the matching of the $\kappa=0.16$ data, for example, depending on various
``cuts'' imposed on the ratios considered, such as the minimum perimeter and
area of the Wilson loops and the maximum relative error, we obtained average
$\DB$ values between $0.36$ and $0.55$. This is somewhat smaller than the
preferred value from the matching of $m_\pi/m_\rho$, indicating again that we
are not yet in the regime of a unique, observable independent \BF.

The static quark potential provides us with another matching condition between
the large and small lattices. We computed finite $t$ approximations to the
static quark potential using time-like Wilson loops $W(\vec{r},t)$ which were
constructed using ``APE''-smeared spatial links~\cite{ape87a,kennedy96b}. On-
and off-axis spatial paths were used with distances $r=n$, $\sqrt{2}n$,
$\sqrt{3}n$, and $\sqrt{5}n$, with $n$ a positive integer. The ``effective''
potentials
\begin{displaymath}
  V(\vec{r},t) = \ln{W(\vec{r},t) \over W(\vec{r},t+1)}
\end{displaymath}
were fitted using a correlated $\chi^2$ procedure with the spatial covariance
estimated using the bootstrap method. The potential on the large lattice was
fitted to the form\footnote{We note that we never observed string breaking.}
\begin{displaymath}
  V(\vec{r}) = V_0 + \sigma r
    - {e\over r} - e'\Bigl(G_L(\vec{r}) - {1\over r}\Bigr);
\end{displaymath}
the last term takes account of the lattice artifacts at short distance. Here,
$G_L(\vec{r})$ denotes the lattice Coulomb potential for the Wilson gluonic
propagator.

\begin{figure}[htb]
  \epsfxsize=0.75\textwidth
  \centerline{\epsfbox{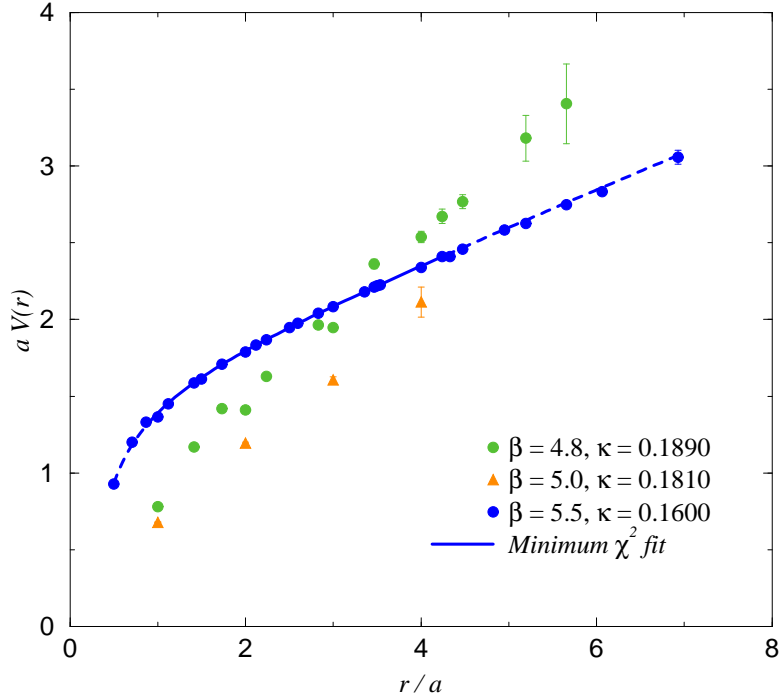}}
  \caption[fig2b]{Non-matching of the $q\bar q$ potential. The line shows the
    ``effective'' potential fitted to the $16^3\times32$ lattice potential at
    $\beta=5.5$, with the solid part of the line indicating the range of $r$
    values used in the fitting procedure. The best fit gives $V_0=0.748(4)$,
    $\sigma=0.0585(6)$, $e=0.340(6)$, and $e'=0.33(1)$ with $\chi^2=8.7$ for
    $14$ degrees of freedom.}
  \label{fig2b}
\end{figure}

In figure~\ref{fig2b}, we plot the potential for the $16^3\times32$ lattice with
$\beta=5.5$ and $\kappa=0.1600$ scaled to a spatial lattice of length $8$. Also
plotted are the potentials for the $8^3\times16$ lattices with $\beta=4.80$ and
$\kappa=0.1890$, and $\beta=5.0$ and $\kappa=0.1810$: the latter point only has
the on-axis values shown. These points have nearly matching pion-rho ratios.

We can see that the $8^3\times16$ potentials do not match the $16^3\times32$
lattice. While we do not expect the constant $V_0$ to match between the fine and
coarse lattices, it is clear that the slopes of the coarse lattice potentials
are too large. The slope decreases with increasing $\beta$, indicating that the
shift $\DB$ required for matching the potential is smaller than that required
for matching $m_\pi/m_\rho$ and closer to that of the Wilson loops ratios. We
also see large violations of rotational symmetry in the coarse lattice potential
compared to the finer lattice potential, indicating that the coarse lattice is
at relatively strong coupling.

\subsection{Parameter reduction at strong coupling}

\begin{figure}[htb]
  \epsfxsize=0.75\textwidth
  \centerline{\epsfbox{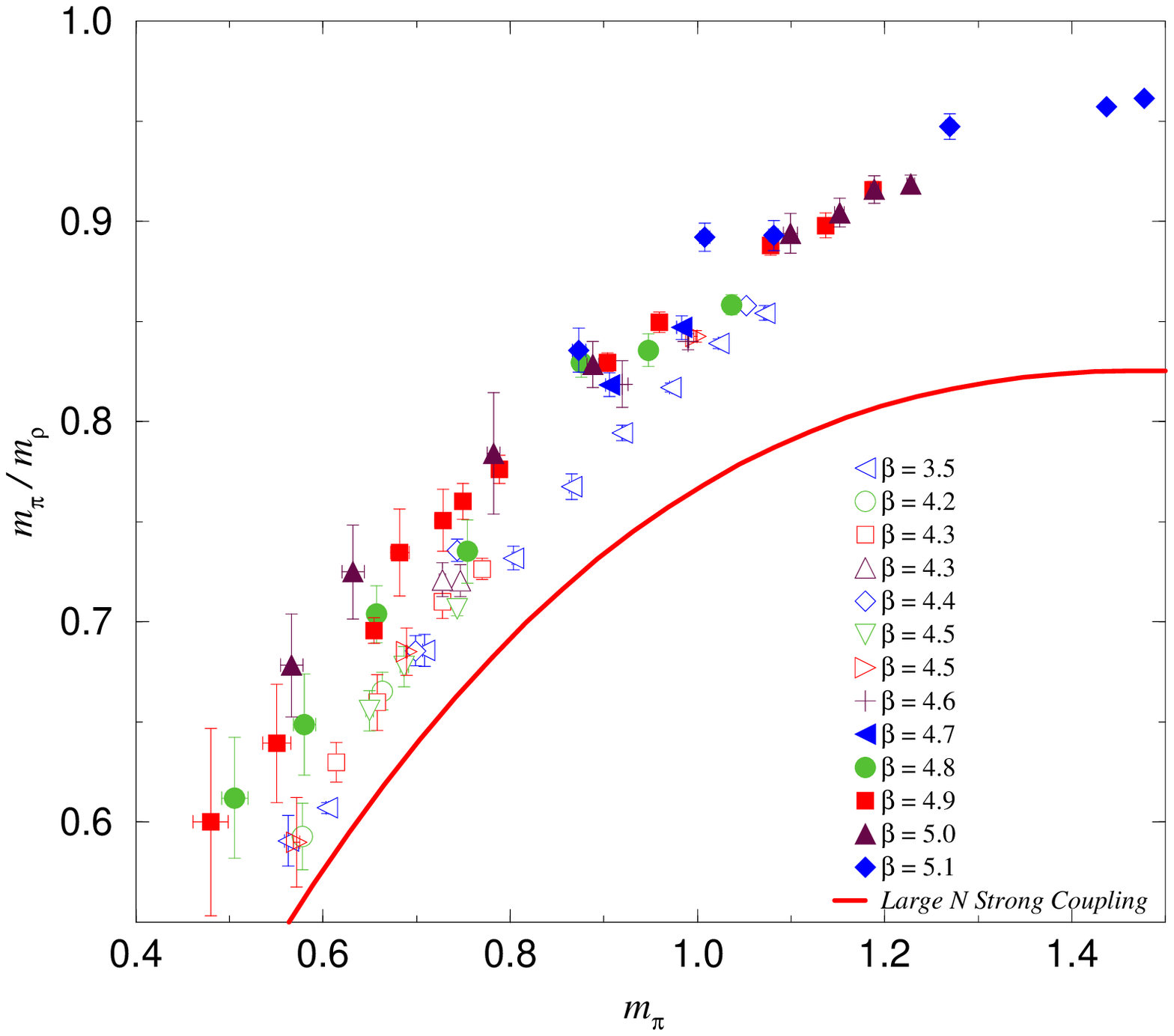}}
  \caption[fig3]{Cumulative data for $3.5\leq\beta\leq5.1$ on an $8^3\times16$
    lattice. The line shows the large $N$ strong coupling result.}
  \label{fig3}
\end{figure}

A third result is an approximate parameter reduction in the theory at strong
coupling ($\beta=3.5$ to~$\beta=4.7$). Figure~\ref{fig3} includes all our data
in this range of $\beta$ on the smaller lattice and shows that for
$3.5\ltapprox\beta\ltapprox4.9$ the data is essentially a function of only a
single parameter. Any change in $\beta$ may be compensated for by a change in
$\kappa$. This is consistent with the large $N$ strong coupling result of
Kawamoto and Smit~\cite{kawamoto81a}, which for $N=3$ and $r=1$ gives
\begin{displaymath}
  m_\pi^2 = 4.8m, \qquad m_\rho = 0.984 + 1.97m.
\end{displaymath}

\section{Conclusions}

\begin{figure}[htb]
  \epsfxsize=0.75\textwidth
  \centerline{\epsfbox{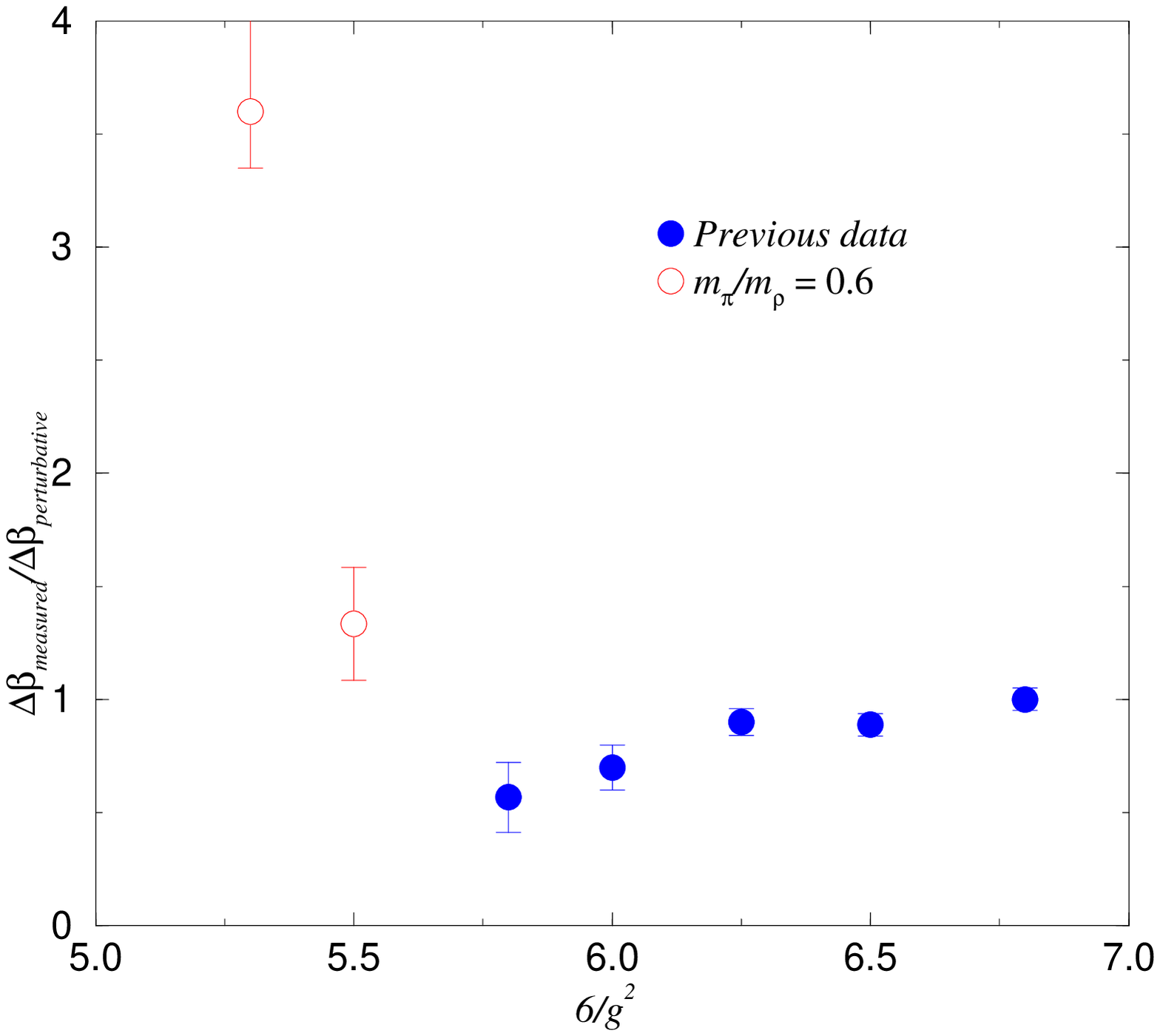}}
  \caption[fig4]{The {\qcd} {\BF} for $m_\pi/m_\rho\approx0.6$. The previous
    data is from reference \protect{\cite{kennedy89a}}.}
  \label{fig4}
\end{figure}

Combining the present results with those obtained earlier on smaller lattices
\cite{bitar89b} we find the {\BF} shown in figure~\ref{fig4}. It is similar to
that found for pure $SU(3)$ lattice gauge theory with the Wilson gauge action.
We are led to the conclusion that current dynamical simulations are being done
at strong coupling.

Indeed, not only do we find that for $\beta\ltapprox5.5$ the {\BF} does not
agree with the perturbative predictions (asymptotic scaling), but that different
renormalization conditions give different estimates for the {\BF}: the shift
$\DB=0.6(1)$ which gives satisfactory matching for $m_\pi$ and $m_\pi/m_\rho$
fails to match $m_\pi/m_N$, ratios of Wilson loops, or the $q\bar q$ potential.
In other words we do not even observe a universal but non-perturbative {\BF}
(scaling).

One therefore may need to reassess estimates for parameters and hence the cost
of future computations with dynamical Wilson fermions accordingly.

\section*{Acknowledgements}
This research was supported by by the U.S. Department of Energy through Contract
Nos. DE-FG05-92ER40742 and DE-FC05-85ER250000.

\bibliographystyle{heller-unsrt}
\bibliography{adk,lattice-bibliography}
\end{document}
\bye